\documentclass[runningheads]{llncs}
\usepackage[T1]{fontenc}
\usepackage{graphicx}
\usepackage{caption}
\usepackage{float}
\usepackage{subcaption}
\usepackage{siunitx}
\begin{document}
\title{Charged particle reconstruction for future high energy colliders with Quantum Approximate Optimization Algorithm}
\titlerunning{Charged particle reconstruction with QAOA}
%
\author{Hideki Okawa\orcidID{0000-0002-2548-6567}}
\authorrunning{Hideki Okawa}
%
\institute{Institute of High Energy Physics, Chinese Academy of Sciences, Shijingshan Beijing 100049, China
\email{okawa@ihep.ac.cn}\\
\url{}}
\maketitle              
\begin{abstract}

Usage of cutting-edge artificial intelligence will be the baseline at future
high energy colliders such as the 
High-Luminosity Large Hadron Collider, to cope with the enormously increasing 
demand of the computing resources. The rapid development of quantum machine learning 
could bring in further paradigm-shifting improvement to this challenge. 
One of the two highest CPU-consuming components, the charged particle reconstruction,
the so-called track reconstruction,
can be considered as a quadratic unconstrained binary optimization (QUBO) problem. The 
Quantum Approximate Optimization Algorithm (QAOA) is one of the most promising algorithms 
to solve such combinatorial problems and to seek for a quantum advantage in the
era of the Noisy Intermediate-Scale Quantum computers.
It is found that the QAOA shows promising performance and demonstrated itself 
as one of the candidates for the track reconstruction using 
quantum computers.

\keywords{Quantum Machine Learning \and QUBO \and QAOA \and High Energy Physics \and track reconstruction }
\end{abstract}
\section{Introduction}

High energy physics aims to unveil the laws of the fundamental building blocks of matter, the elementary particles, 
and their interactions. High energy colliders have been one of the most promising approaches to discover new 
particles and deepen understanding of the underlying physics through precise measurements. After the 
revolutionary discovery of the Higgs boson in 2012~\cite{ATLAS:2012yve,CMS:2012qbp} 
at the ATLAS~\cite{ATLAS:2008xda} and CMS~\cite{CMS:2008xjf} experiments 
at the Large Hadron Collider (LHC)~\cite{Evans:2008zzb}, 
we are entering the precision measurement era of the Higgs sector, first of all, to be pursued at the 
High-Luminosity LHC (HL-LHC)~\cite{hl-lhc}, and to be followed by future colliders being proposed, such as 
the Circular Electron Positron Collider (CEPC)~\cite{CEPC-SPPC_preCDR1,CEPC-SPPC_preCDR2,CEPC_TDR1,CEPC_TDR2} to be hosted in China. 

At the HL-LHC, we will enter the “Exa-byte” era, where the annual computing cost will increase by a factor of 10 to 20. Without various innovations, the experiment will not be able to operate. Usage of the Graphical Processing Units (GPUs) and other state-of-the-art artificial intelligence (AI) technologies such as deep learning will be the baseline at the HL-LHC. However, the emerging rapid development of quantum computing and implementation of machine learning techniques in such computers could bring in another leap.
Two of the highly CPU consuming components at the LHC and HL-LHC are (1) the charged particle reconstruction, the so-called track reconstruction,  for both in data and simulation and (2) simulation of electromagnetic and hadronic shower development in the calorimeter. Development of quantum machine learning techniques to overcome such challenges would not only be important for the HL-LHC, but also for a future Higgs factory CEPC and a next generation discovery machine the Super Proton-Proton Collider (SppC)~\cite{CEPC-SPPC_preCDR1,CEPC-SPPC_preCDR2} to be hosted in China as well as other such colliders under consideration in the world.


\section{Track Reconstruction}

Track reconstruction or tracking is the standard procedure in the collider experiments to identify charged particles 
traversing the detector and to measure their momenta. The curvature of particle trajectories (tracks)
bent in a magnetic field will provide the momentum information.
Tracks are reconstructed from hits in the silicon detectors, which have 
many irrelevant hits from secondary particles and detector noise, and require sophisticated algorithms. 
Tracking is one of the most crucial components of reconstruction in the collider experiments.

At the HL-LHC, additional proton-proton interactions per bunch crossing, the so-called pileup, becomes 
exceedingly high, and the CPU time required to run the track reconstruction 
explodes with pileup~\cite{ATL-PHYS-PUB-2019-041,Cerati:1966040}.

\subsection{Current classical benchmarks}

The Kalman Filter technique~\cite{Fruhwirth:1987fm} 
has been often used as a standard algorithm to reconstruct the tracks.
It is implemented in A Common Tracking Software (ACTS)~\cite{acts}, for example. 
Seeding from the inner detector layers, the tracks are extrapolated to predict the next hit and 
iterated to find the best quality combination. It is a well established procedure and 
has excellent performance but suffers 
from the computing time, especially when the track multiplicity per event becomes high. 

Recently, usage of the graph neural network (GNN) is actively investigated at the LHC~\cite{Ju:2020xty,Lazar:2022ixi} 
as well as other collider experiment including the tau-lepton and charm-quark factory 
in China: Beijing Spectrometer (BES) III~\cite{BESIII:2009fln}. Hits in the silicon detectors can be regarded as 
“nodes” of the graphs, and segments reconstructed by connecting the hits can be 
considered as “edges”. GNN-based algorithms provide compatible performance as the 
Kalman Filter, but the computing time scales approximately linearly instead of exponentially with the 
number of tracks. The GNN is thus considered to be one of the new standards in the era of 
the HL-LHC. 

\subsection{Quantum approach}

There have been several studies to run the track reconstruction with quantum computers. 
First of all, doublets are formed by connecting two hits in the silicon detectors. Then, triplets, 
segments with three silicon hits, are formed by connecting the doublets. 
Then, triplets are connected to reconstruct the tracks, by evaluating the consistency
of the triplet momenta. Such procedure can be considered as a quadratic unconstrained binary 
optimization (QUBO) problem:
\begin{equation}
	O(a,b,T) = \sum_{i=1}^{N} a_i T_i + \sum_{i=1}^{N} \sum_{j<i}^{N} b_{ij} T_i T_j,
	\label{eq:qubo}
\end{equation}
where $N$ is the number of triplets, $T_i$ and $T_j$ corresponds to the triplets and  
takes the value of either zero or one,
$a_i$ are the bias weights to evaluate the quality
of the triplets, and $b_{ij}$ are the coefficients that quantify the compatibility of 
two triplets ($b_{ij}=0$ if no shared hit, $=1$ if there is any conflict, and $=-S_{ij}$ 
if two hits are shared between the triplets). The coefficients $-S_{ij}$ quantify 
the consistency of the two triplet momenta by~\cite{Bapst:2019llh}:
\begin{equation}
	S_{ij} = \frac{1-\frac{1}{2}(|\delta(q/p_{Ti}, q/p_{Tj})| + max(\delta\theta_i,\delta\theta_j))}{(1+H_i+H_j)^2},
\end{equation}
where $\delta$ is the difference between the curvature $q/p_{T}$ or angle $\theta$ of the two triplets
and $H_i$ is the number of holes in the triplet. 

The bias weights $a_i$ have significant impact on the Hamiltonian 
energy landscape and thus on the track reconstruction and computation speed. They can be parameterized
as: 
\begin{equation}
	a_i = \alpha \left(1 - e^{\frac{\mid d_0 \mid}{\gamma}} \right) + \beta \left(1 - e^{\frac{\mid z_0 \mid}{\lambda}} \right), 
\end{equation}
where $d_0$ and $z_0$ are the transverse and longitudinal displacements of the triplets from 
the primary vertex (the most significant proton-proton collision point in an event)
and $\alpha$, $\beta$, $\gamma$ and $\lambda$ are tunable parameters, which 
are taken to be 0.5, 0.2, 1.0 and 0.5, optimized in a previous study~\cite{lucy} for the same 
dataset (see Section~\ref{sec:dataset}). 

The first quantum tracking studies~\cite{Bapst:2019llh,Zlokapa:2019tkn} 
have been pursued with quantum annealing computers. 
The quantum annealer looks for the global minimum of a given function with quantum 
tunneling. It is a natural machine to solve QUBO problems by searching for the 
ground state of a Hamiltonian. As the size of the QUBO generally exceeds the number of available 
qubits in the current era of the Noisy Intermediate-Scale Quantum (NISQ) 
computers~\cite{Preskill2018quantumcomputingin}, 
the QUBO is split into subsets (sub-QUBOs) to search for the ground state. 
Performance evaluation between the D-Wave annealing machine and a classical emulation
using a digital annealer is mentioned in Ref.~\cite{Saito:2020pld}.
The performance dependence against the size of the sub-QUBOs is evaluated in Ref.~\cite{Schwagerl:2023elf}.

To exploit the quantum gate computers, 
a QUBO can be mapped to an Ising Hamiltonian and be solved using the Variational Quantum 
Eigensolver (VQE), Quantum Approximate Optimization Algorithm (QAOA)~\cite{farhi2014quantum},
or other similar algorithms.
Previous studies conducted at the LUXE experiment considered the TwoLocal ansatz with the R$_{\mathrm{Y}}$ gates 
and a circular CNOT entangling pattern~\cite{Funcke:2022dws,Crippa:2023ieq}. 
The LHCb Collaboration has also looked into the QUBO approach with the gate computer using 
the Harrow-Hassadim-Lloyd (HHL) algorithm~\cite{Nicotra:2023rmn}.

Another distinctive approach is pursued with the hybrid quantum classical GNN,
where some components of the classical GNN method~\cite{Ju:2020xty,Lazar:2022ixi} are 
replaced by the Variational Quantum Layers. This hybrid model performs similarly to the 
classical approach, as also confirmed by Refs.~\cite{Funcke:2022dws,Crippa:2023ieq}.
This hybrid quantum-classical GNN approach is out of the scope of this paper, and will 
not be mentioned further. 

\section{Datasets}
\label{sec:dataset}

In this study, the dataset from the TrackML Challenge is adopted~\cite{Amrouche:2019wmx,Amrouche:2021nbs}.
It is an open source dataset representing the conditions for the ATLAS and CMS
experiments at the HL-LHC. 
It assumes the particle multiplicities expected with the pileup condition of 
$\left<\mu\right>$=200. The dataset is simplified by focusing on the barrel (central)
region of the detector, thus removing the hits in the end-cap region, 
which is a common approach often adopted for tracking studies for the HL-LHC. 

The QUBO matrix, namely the bias weights $a_i$ and the compatibility coefficients $b_{ij}$ 
as defined in Equation~\ref{eq:qubo}
is extracted from the dataset using the hepqpr-qallase library~\cite{hepqpr}.
The QUBO matrix is pretty sparse, as is the nature of the collider experiments and
track reconstruction. 

\section{Methodology}

\begin{figure}[t]
    \centering
        \includegraphics[width=\linewidth]{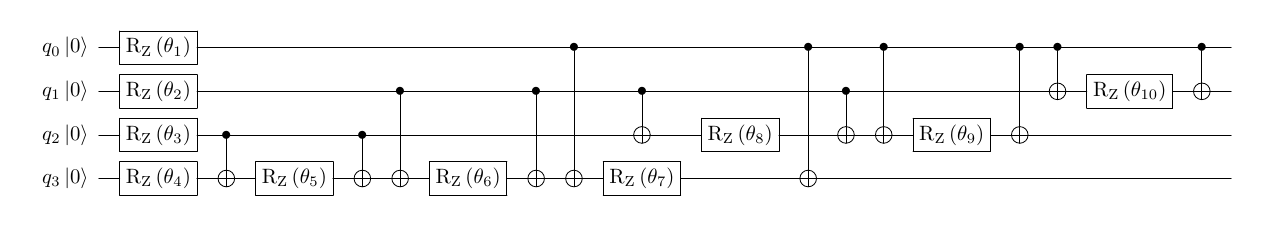}
    \caption{Layout of the QAOA with the Quantum Alternative Operator Ansatz. The number of qubits is reduced to four for the demonstration purpose.}
    \label{fig:circuit}
\end{figure}

The QAOA is considered in this study.
It is a hybrid quantum-classical method, designed to solve combinatorial 
optimization problems. 
Even at the lowest circuit depth, the QAOA cannot efficiently be simulated by classical 
computers, but has non-trivial guarantees on the performance~\cite{farhi2014quantum,farhi2015quantum}. 
For cases where the minimum energy spectrum gap is very small, the computing time 
required in the quantum annealers is very long to remain adiabatic. The QAOA is 
known to outperform adiabatic quantum annealing by several orders of magnitude in such 
circumstances~\cite{PhysRevX.10.021067}.
Thus, the QAOA is one of the most promising algorithm to seek for quantum advantage in the 
era of the NISQ computers. 

Its libraries are implemented in pyqpanda-algorithm~\cite{pyqpanda} by Origin Quantum
(Chinese name: Benyuan).
It adopts the Quantum Alternative Operator Ansatz~\cite{qaoaAnsatz}. The schematic layout of the 
quantum circuit is demonstrated in Fig.~\ref{fig:circuit}.
For the actual hardware computation, the 6-qubit machine Wuyuan is 
used through the cloud service~\cite{benyuan}.

\subsection{Optimization of QAOA implementation}

\begin{figure}[p]
    \centering
    \begin{subfigure}{\linewidth}
        \includegraphics[width=\linewidth]{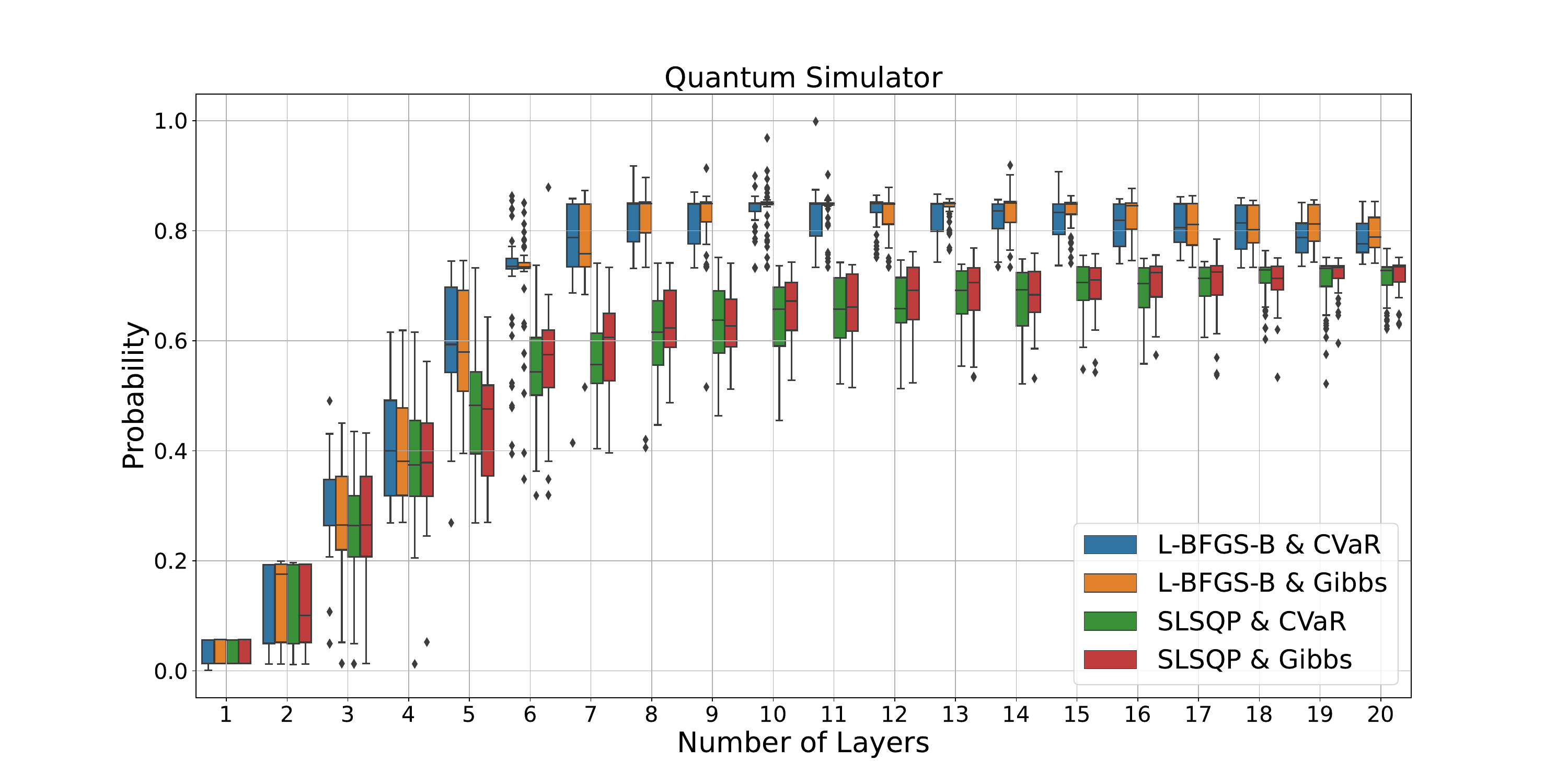}
    \caption{}
    \label{subfig:qaoa6}
    \end{subfigure}
    \begin{subfigure}{\linewidth}
        \includegraphics[width=\linewidth]{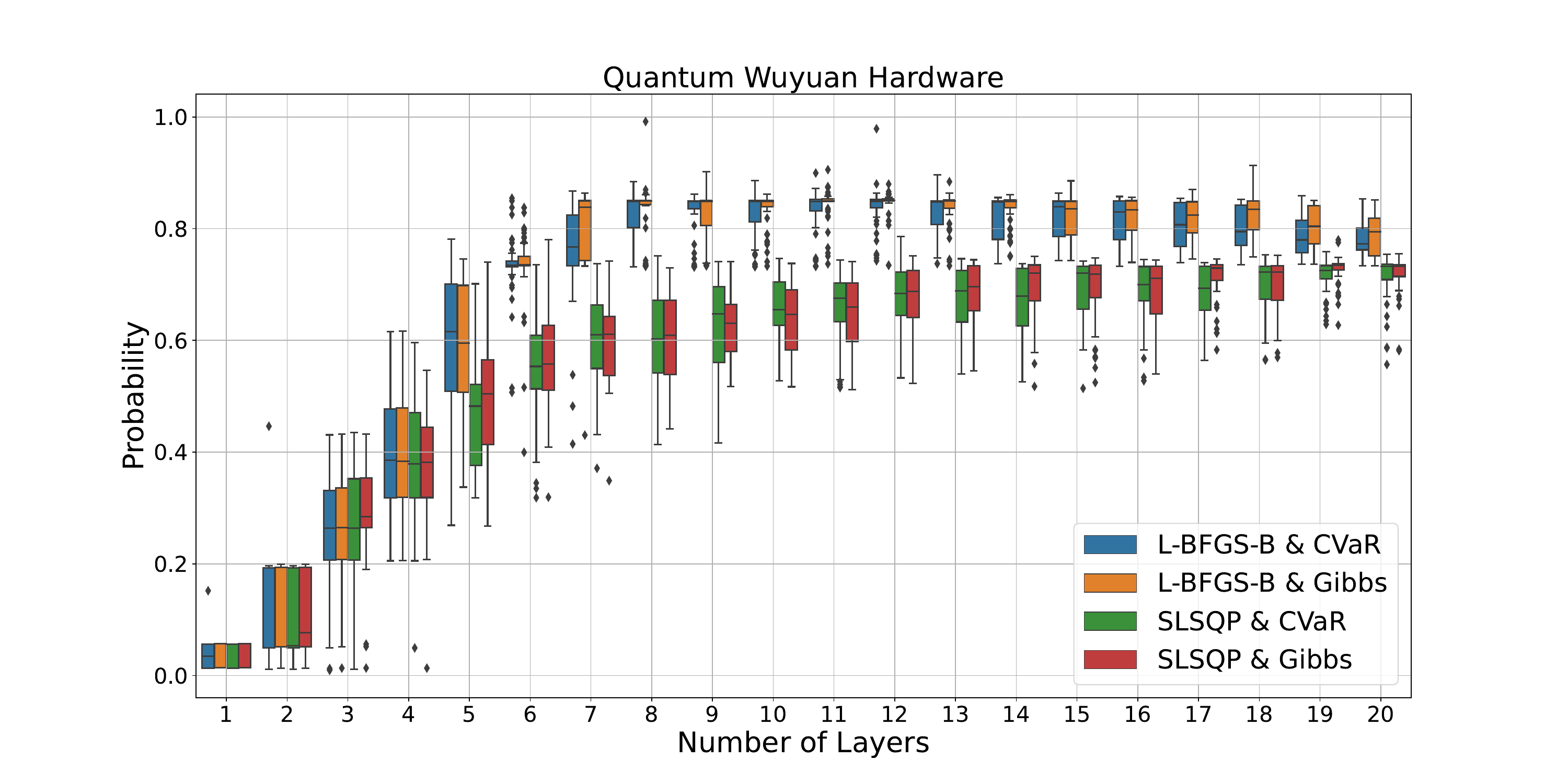}
    \caption{}
    \label{subfig:qaoa6_real}
    \end{subfigure}
    \caption{Probability of finding the correct combination with the L-BFGS-B or SLSQP optimizers and the CVaR or Gibbs
        loss functions, presented against the number of QAOA layers for the quantum simulator (a) and Wuyuan hardware (b).}
        \label{fig:qaoa6}
\end{figure}

To optimize the conditions to run the QAOA, a 6$\times$6 matrix is extracted 
from a TrackML QUBO to match the available number of qubits in Wuyuan. 
Performance is evaluated for two loss functions: 
CVaR~\cite{Barkoutsos2020improving} and Gibbs~\cite{PhysRevResearch.2.023074}; three optimizers
which can handle bounds on the variables: L-BFGS-B~\cite{byrd1995limited,L-BFGS-B}, 
Sequential Least SQuares Programming (SLSQP)~\cite{SLSQP}, 
and a Truncated Newtonian algorithm (TNC)~\cite{byrd1995limited};
and the number of layers of the QAOA. 

Fig.~\ref{fig:qaoa6} shows the probability of finding the correct 
minimum energy with various loss functions, optimizers and the number
of the QAOA layers for the quantum simulator and the actual quantum hardware. 
The probabilities tend to be low for the shallow QAOA, 
which is consistent with what is reported in 
Refs.~\cite{PhysRevLett.125.260505,hastings2019classical,farhi2020quantum,farhi2020quantum2}.
Among the three optimizers, L-BFGS-B shows the best performance, 
followed by SLSQP. TNC shows largely degraded probabilities
without much improvement against the number of layers, thus, is not 
presented in the figure. There is no significant difference between
the CVaR and Gibbs loss functions, thus CVaR is adopted onward in this work. 
The real hardware shows compatible
performance as the quantum simulator, and there is no sign of 
degradation even for deep-layer QAOAs up to 20 layers. This is 
in contrast to what is observed in Ref.~\cite{noisyQAOA} but consistent
with Ref.~\cite{Schwagerl:2023elf}. This could 
be due to the difference in the QUBO considered, 
which is largely sparse in this paper as well as in Ref.~\cite{Schwagerl:2023elf}. 

\begin{figure}[t]
    \centering
    \begin{subfigure}{0.49\linewidth}
        \includegraphics[width=\linewidth]{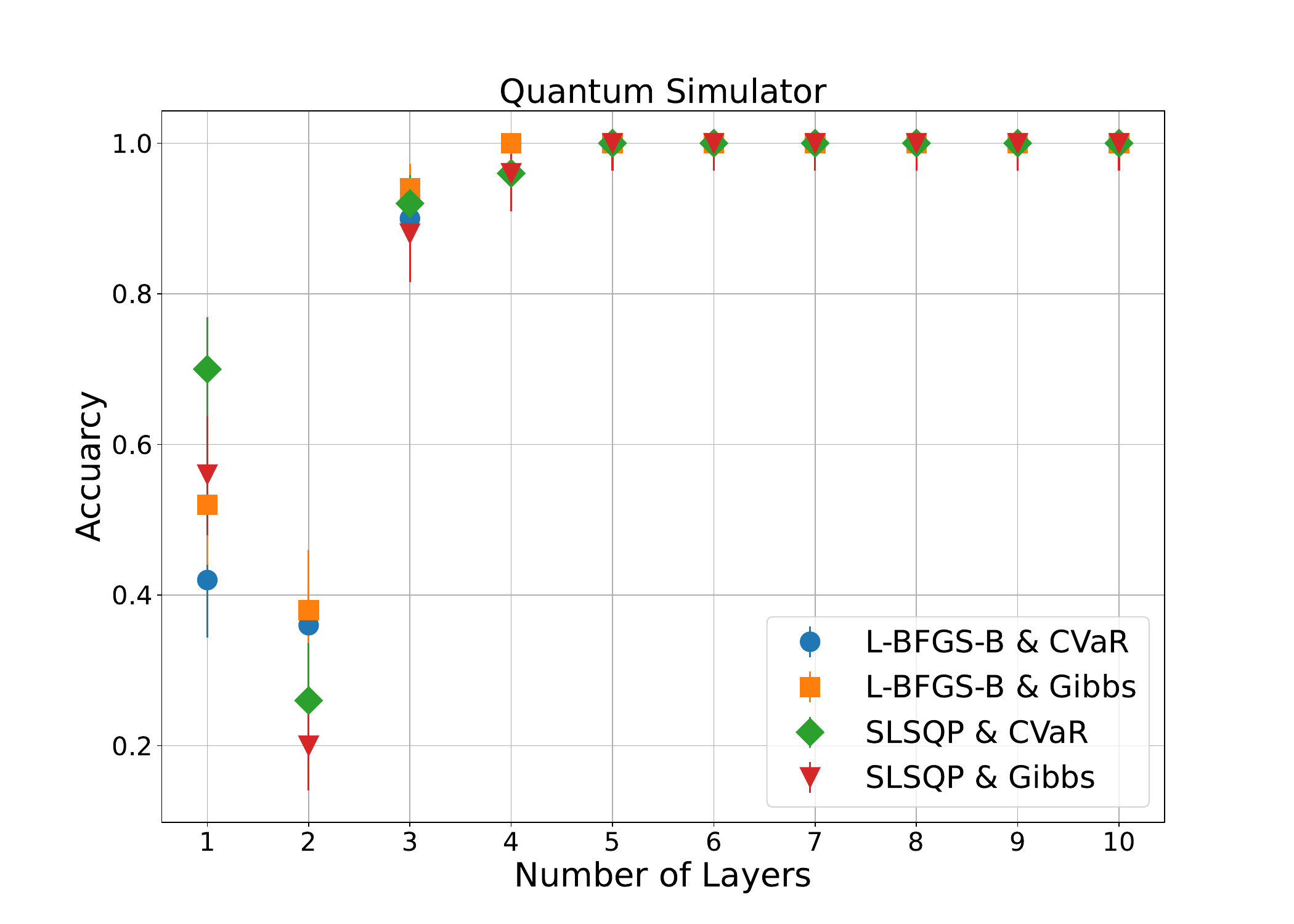}
    \caption{}
    \label{subfig:qaoa6}
    \end{subfigure}
    \begin{subfigure}{0.49\linewidth}
        \includegraphics[width=\linewidth]{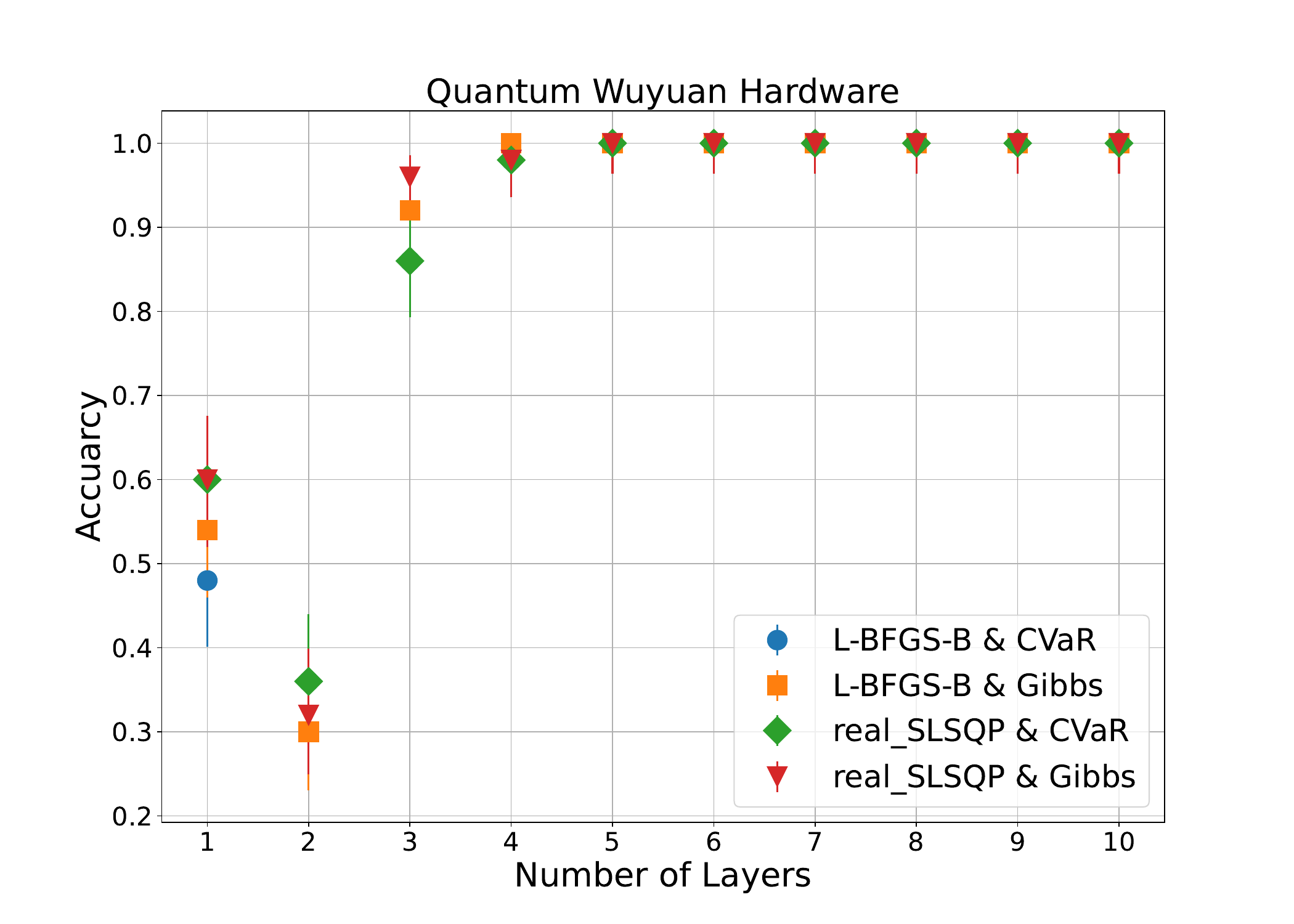}
    \caption{}
    \label{subfig:qaoa6_real}
    \end{subfigure}
	\caption{Accuracy of finding the correct combination in a single job with the L-BFGS-B or SLSQP optimizers and the CVaR or Gibbs loss functions, presented against the number of QAOA layers for the quantum simulator (a) and real hardware (b). The statistical uncertainty is presented in the error bars.}
        \label{fig:qaoa6_acc}
\end{figure}

It is worth emphasizing that in the actual implementation, a
single QAOA job will run multiple shots and the combination with the
highest probability will be selected.
Thus, the accuracy of obtaining the correct answer is much higher
than the probability itself, reaching 100\% already at 5 layers 
as is presented in Fig.~\ref{fig:qaoa6_acc}.
In the following sections, seven QAOA layers are adopted, where the probability
reaches the plateau and the accuracy is compatible with 100\% within the 
statistical uncertainty.

\subsection{Sub-QUBO method}

The number of triplet candidates define the number of qubits required. 
Obviously the quantum computing resources currently available in the NISQ era 
cannot cover the full QUBO for tracking. Thus, the QUBO is split into sub-QUBOs of 
size 6$\times$6 to match the Wuyuan hardware. 

There are various sub-QUBO algorithms proposed: qbsolv~\cite{qbsolv} (now in the dwave-hybrid library), 
for example. In this paper, a sub-QUBO method using multiple solution instances~\cite{subQUBO} 
is adopted. This method has a strong theoretical justification, whereas other existing 
approaches are heuristic and lack in such a foundation~\cite{subQUBO}. 
In this multiple solution instance method, 
three parameters ($N_I$, $N_E$, $N_S$) are considered. 
First of all, $N_I$ quasi-optimal solutions are extracted from the full-QUBO classically.
Then $N_S$ solution instances from $N_I$ are randomly selected.
The method focuses on a particular binary variable $T_i$ (see Equation~\ref{eq:qubo}), 
rank them in accordance to how 
much they vary over $N_S$ solution instances. Highly varying $T_i$ will be included in the sub-QUBO model.
The pick-up process of $N_S$ solutions from the QAOA is repeated $N_E$ times and $N_E$ sub-QUBO 
models are considered. Finally, a pool of $N_I$ solutions is returned and the best solution 
is chosen.

\begin{figure}[t]
    \centering
	\begin{subfigure}{\linewidth}
        \includegraphics[width=\linewidth]{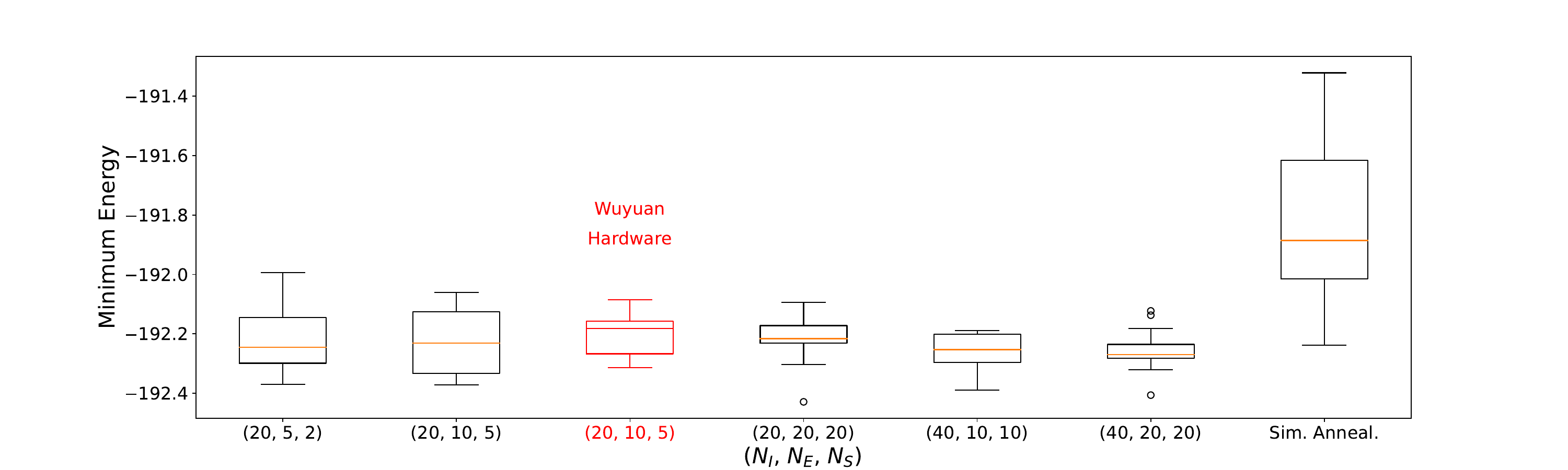}
    	\caption{}
    	\label{subfig:subQUBO_0p05}
    	\end{subfigure}
	\begin{subfigure}{\linewidth}
        \includegraphics[width=\linewidth]{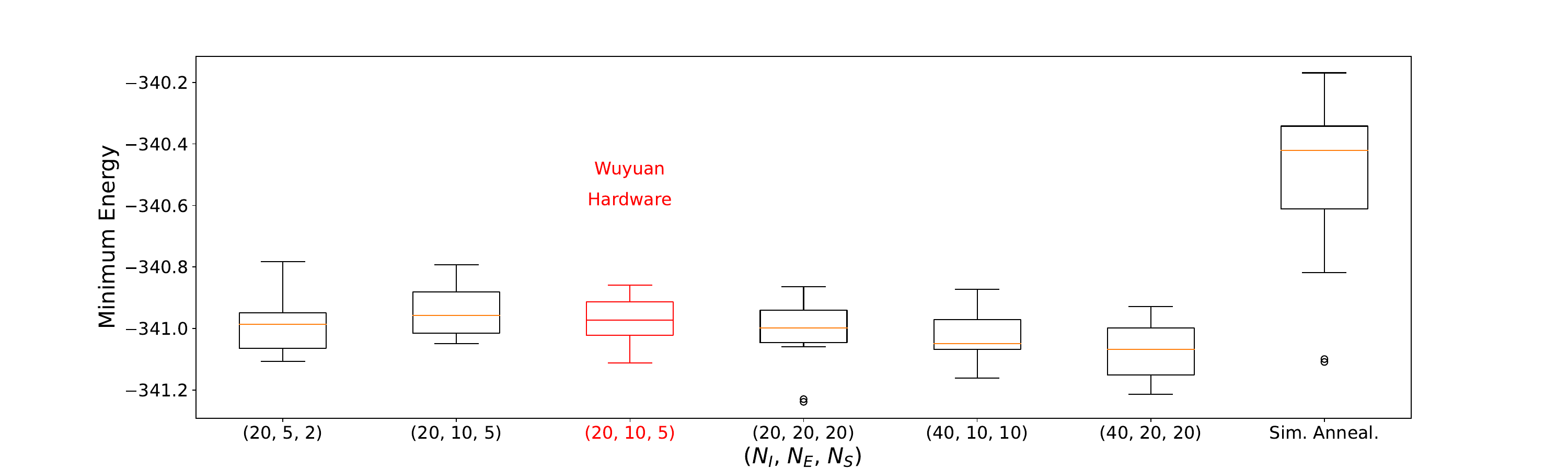}
        \caption{}
        \label{subfig:subQUBO_0p1}
        \end{subfigure}
	\caption{Minimum energy estimated by the multiple solution instance method for various $(N_I, N_E, N_S)$ parameters with the quantum simulator and Wuyuan hardware compared to a classical simulated thermal annealing. Two examples with the full QUBO size of 778$\times$778 (a) and 1431$\times$1431 (b) are presented.}
        \label{fig:subQUBO}
\end{figure}

Fig.~\ref{fig:subQUBO} shows the presumed minimum energy found by 
the multiple solution instance method for various sets of ($N_I$, $N_E$, $N_S$) 
parameters with the quantum simulator compared
to a simple classical optimizer with the simulated thermal annealing. 
Those two examples have the full QUBO size of 778$\times$778 and 1431$\times$1431
respectively, and correspond to the first two points later presented in 
Fig.~\ref{fig:track_perf} in Section~\ref{sec:results}.
The output from the real quantum hardware Wuyuan is also 
presented for one set of the ($N_I$, $N_E$, $N_S$) parameters. 
It is clearly observed that this sub-QUBO method using the QAOA 
succeeds in obtaining lower energy than the classical simulated
annealing. 
There is no obvious dependence in performance on the 
three parameters. These two features of the method are consistent
with what have been reported in the original proposal of this 
sub-QUBO method~\cite{subQUBO}.
As the computing time increases with the size of the parameters, 
$(N_I, N_E, N_S)=(20,10,5)$ is adopted for the final evaluation 
on the tracking performance, which will be summarized in the next section.
It is also promising to see that there is no 
degradation in performance with the actual quantum hardware 
despite the presence of noise. 

\section{Results and Discussions}
\label{sec:results}

Several events are selected containing a few thousands particles and noise. 
The track reconstruction is pursued by running the sub-QUBO method with 
the three parameters $(N_I, N_E, N_S)$ defined in the previous section. The 
QAOA utilizes seven 
layers and the CVaR loss function, and split into sub-QUBOs with the size 
of six qubits. 

Performance of the track reconstruction is evaluated in terms of efficiency
(recall) and purity (precision). They are defined as the following: 
\begin{equation}
	\mathrm{Efficiency} = \frac{TP}{TP+FN} 
	= \frac{\mathrm{\#~of~matched~reconstructed~doublets}}{\mathrm{\#~of~true~doublets}},
\end{equation}
\begin{equation}
        \mathrm{Purity} = \frac{TP}{TP+FP} 
	= \frac{\mathrm{\#~of~matched~reconstructed~doublets}}{\mathrm{\#~of~all~reconstructed~doublets}},
\end{equation}
where $TP$ is the true positives, $FN$ the false negatives, and $FP$
the false positives. $TP$ corresponds to the number of reconstructed
doublets matching to the correct (true) doublets. $FN$ is the 
number of true doublets that are not reconstructed, thus $TP+FN$ 
is simply the number of true doublets. $FP$ is the number of 
reconstructed doublets that do not match to the true doublets. 
In the high energy physics terminology, they are called the ``fake''
doublets. 

\begin{figure}[t]
    \centering
    \begin{subfigure}{0.49\linewidth}
        \includegraphics[width=\linewidth]{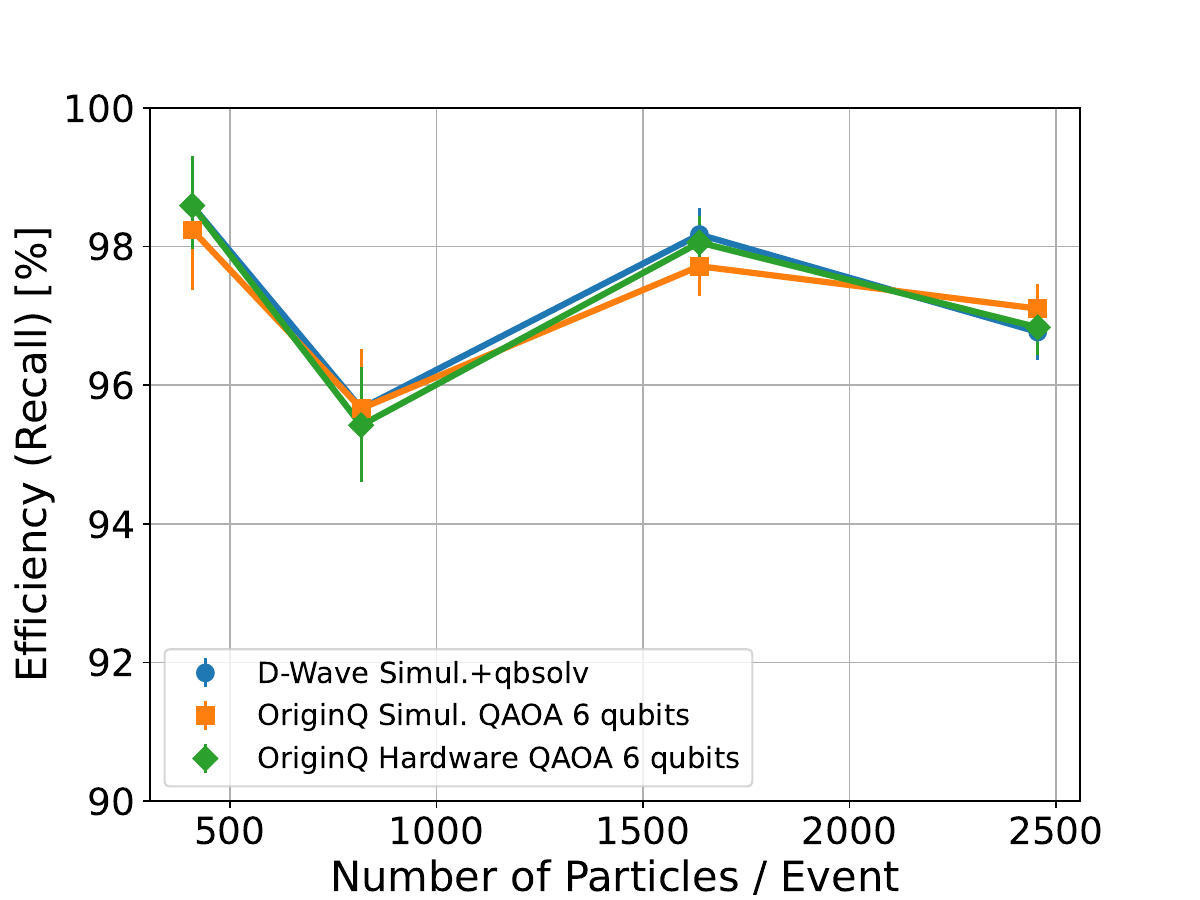}
    \caption{}
    \label{subfig:qaoa6}
    \end{subfigure}
    \begin{subfigure}{0.49\linewidth}
        \includegraphics[width=\linewidth]{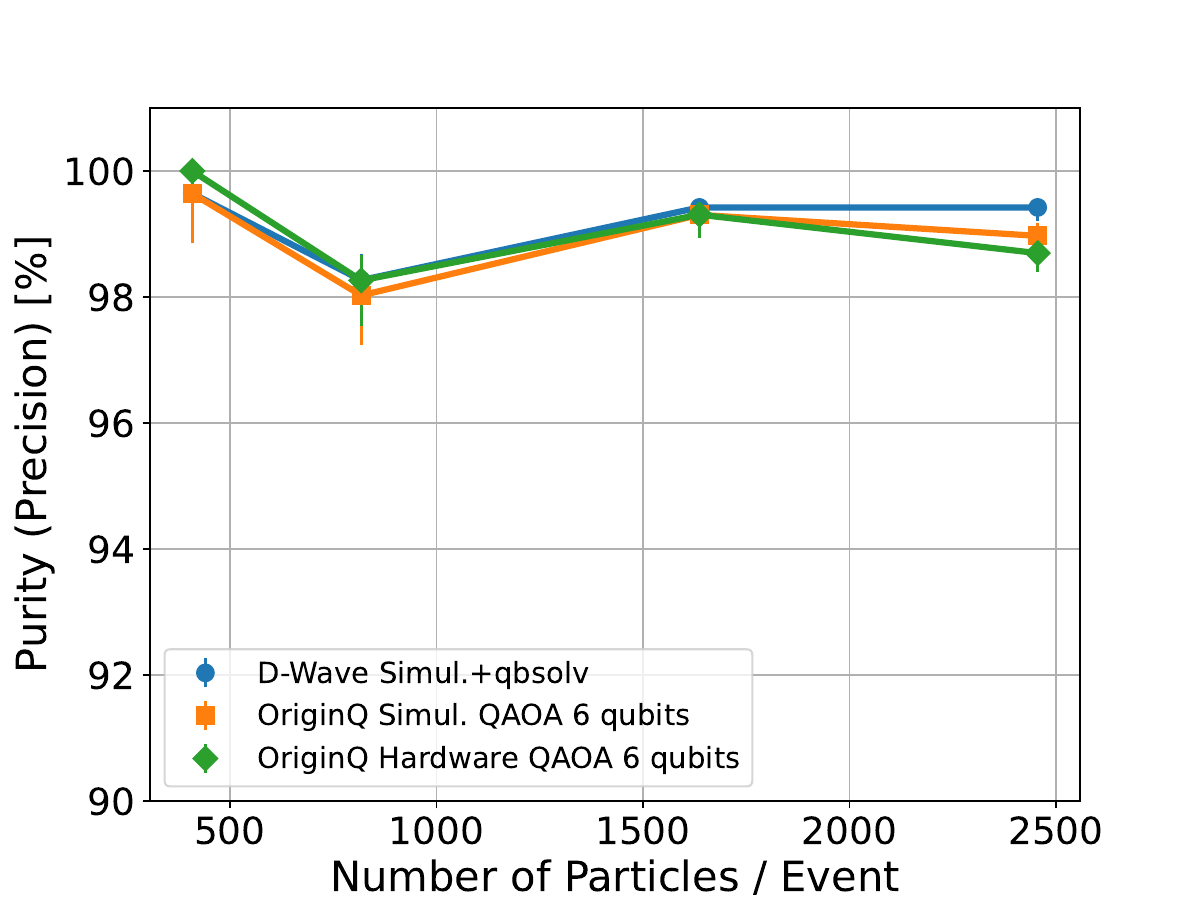}
    \caption{}
    \label{subfig:qaoa6_real}
    \end{subfigure}
	\caption{Efficiency (a) and purity (b) as a function of particle multiplicity utilizing the QAOA simulator, Wuyuan hardware or the D-Wave simulator with qbsolv. }
        \label{fig:track_perf}
\end{figure}

\begin{figure}[p]
    \centering
    \begin{subfigure}{\linewidth}
        \includegraphics[width=0.9\linewidth]{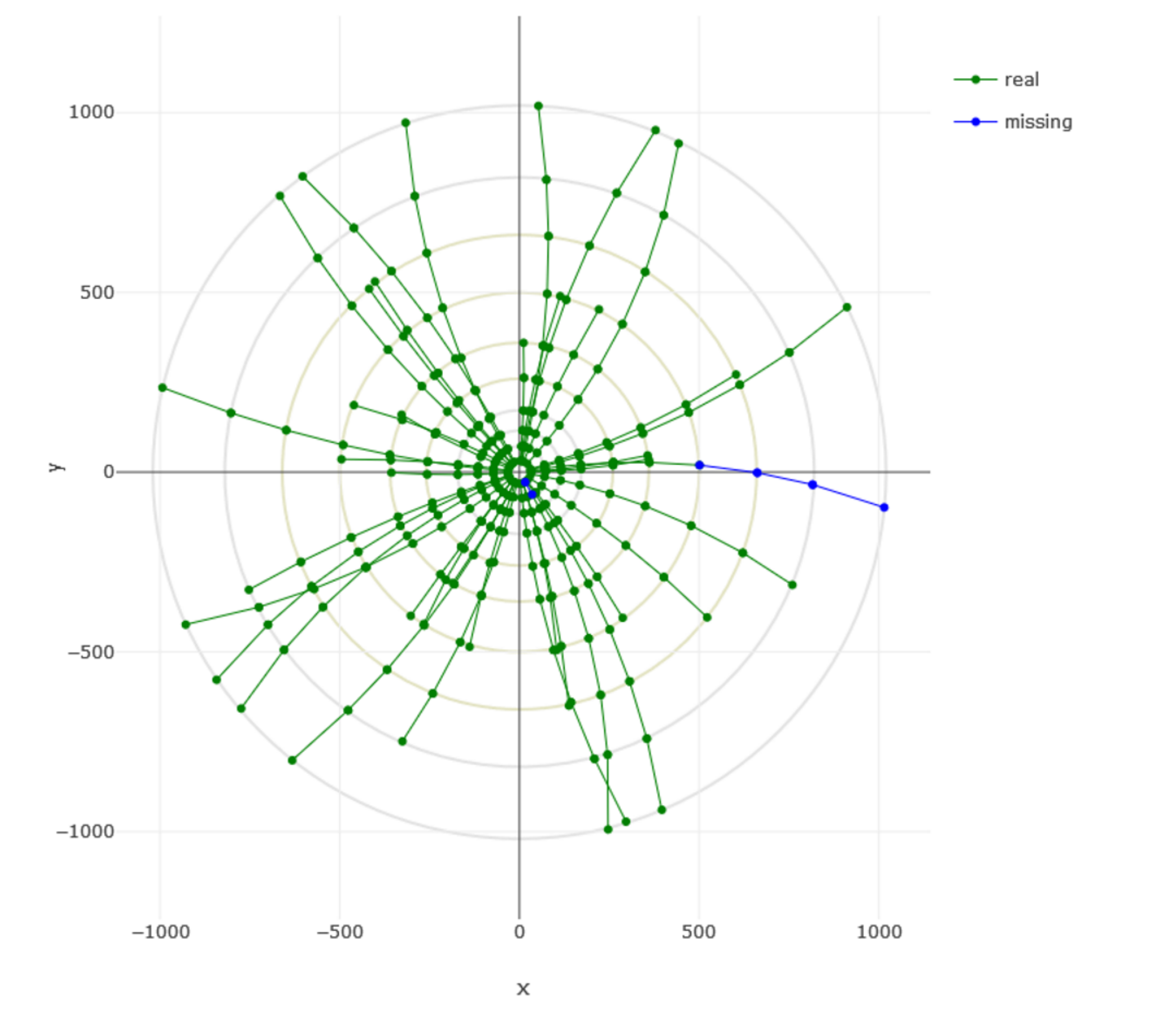}
    \caption{}
    \label{subfig:display_0p05}
    \end{subfigure}
    \begin{subfigure}{\linewidth}
        \includegraphics[width=0.9\linewidth]{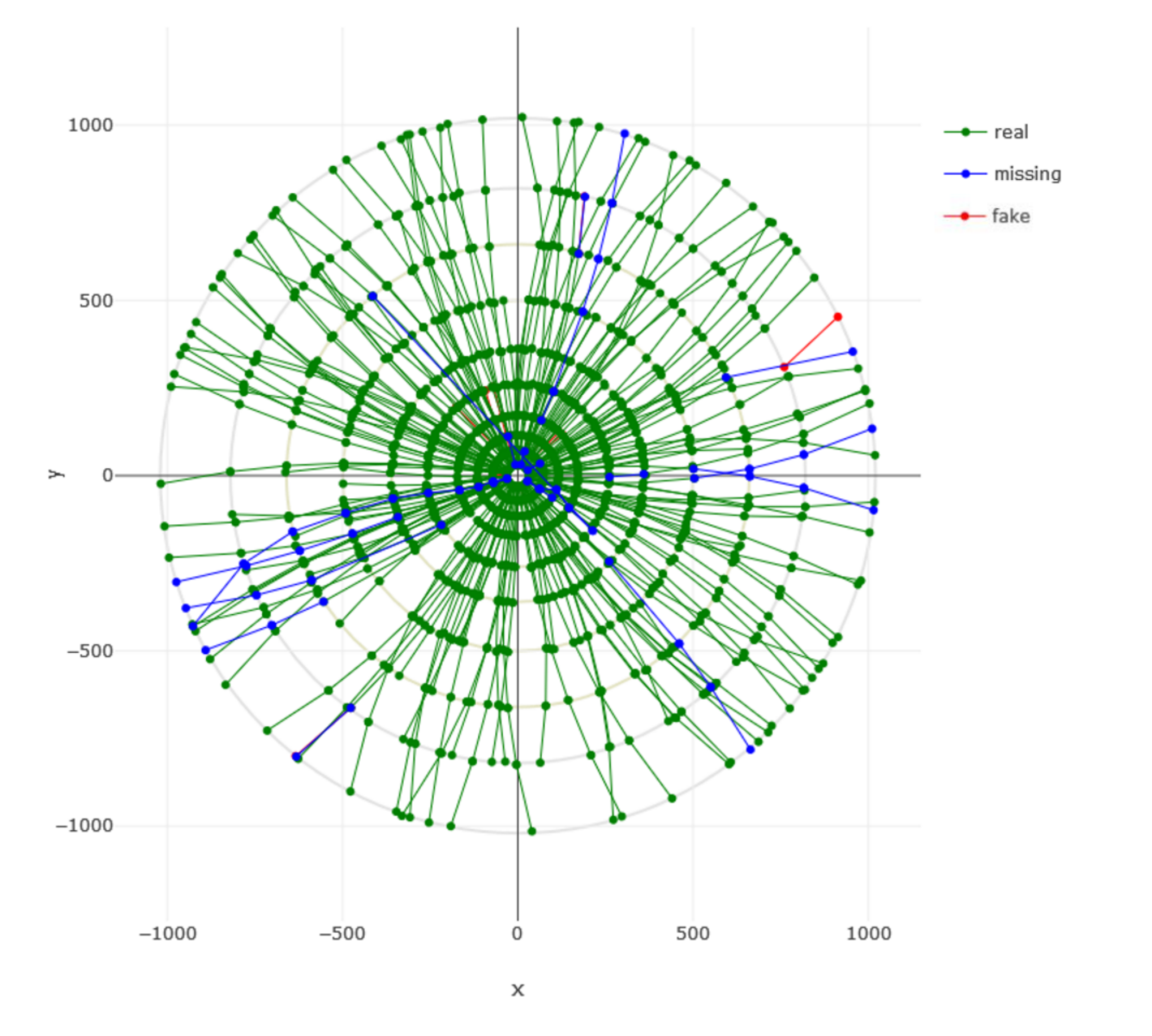}
    \caption{}
    \label{subfig:display_0p2}
    \end{subfigure}
	\caption{Event displays of tracks reconstructed with the QAOA utilizing the Wuyuan hardware. They are generated with hepqpr-qallase library~\cite{hepqpr}.}
        \label{fig:display}
\end{figure}

Fig.~\ref{fig:track_perf} shows the track efficiency and purity 
using the QAOA with the quantum simulator or Wuyuan hardware. The 
performance is compared to the D-Wave simulator approach implemented
in Ref.~\cite{Bapst:2019llh}. The D-Wave simulator with qbsolv 
and NEAL show almost indistinguishable results, so only the qbsolv
values are shown in the figure. 
The track reconstruction performance with QAOA 
is compatible with the D-Wave annealing approach, and there is no 
sign of degradation in the actual hardware. 
Fig.~\ref{fig:display} presents two event displays from the lowest 
and highest particle density events considered in Fig.~\ref{fig:track_perf}.

This work demonstrates that the QAOA is a promising candidate to 
be considered for the track reconstruction at the future colliders. 
With utilizing a robust sub-QUBO method and the QAOA, high 
performance can 
be achieved even with a small size of qubits. The compatible 
performance obtained with the real quantum hardware further 
supports its potential usability. 

It is not yet at the stage to evaluate the quantum advantage, 
since the QAOA with six qubits can run quickly on the quantum 
simulator as well with compatible performance. The non-trivial 
impact is to be investigated 
with higher qubit conditions, which would be beyond the reach of the
quantum simulator and is left for future studies. 

\subsubsection{Acknowledgements} The author would like to thank 
Federico Meloni and David Spataro for discussions regarding 
the quantum tracking and Andreas Salzburger for his suggestion on 
the TrackML dataset. The author would also like to thank Ziwei 
Cui and Lei Li from Origin Quantum (Benyuan) for various feedback. 
The author is supported by NSFC under contract No.~12075060.
This work is benefited by the libraries and quantum computing
resources provided by Origin Quantum.

\bibliographystyle{splncs04}
\bibliography{qtrack}

\end{document}